\documentclass[referee]{aa}
\usepackage{graphics}
\begin{document}
\title{Gravitational Shapiro Telescope on the period of pulsars
to discover Dark Planets and MACHOs}
\author{D. Fargion \inst{1} \inst{2} \and R. Conversano \inst{3}}
\institute{INFN, Rome, Italy \and Phys. Dept. University of Rome
``La Sapienza'', Rome, Italy \and HPCN - ENEA, Rome, Italy}
\date{Received / Accepted}
\maketitle
\begin{abstract}
Collective Shapiro Phase Shift on the period of the pulsars due to
a dark object (a passing MACHO or a planet) along the
line-of-sight of one or more pulsar is a formidable gravitational
tool to discover dark matter.\\ We already noted that the presence
of a few negative pulsar periods, possibly  due to such a
Shapiro delay, is roughly consistent with independent estimates of
MACHOs observed by microlensing in our Galaxy. Here we update our
study and suggest to verify and to calibrate the Shapiro phase
delay to observe on ecliptic  main known planetary and solar
gravitational fields. Once the test is probed we propose to use,
in a collective way, a subsample of pulsars on the ecliptic plane to
monitor our solar system for discovering any heavy unknown object,
either dark MACHOs or unknown planets. The importance of such
unknown possible planetary components and their eventual near
encounter with Earth in the far past (or future) should be not
neglected : the  same life evolution (or extinction) might be
marked by such a rare event. \keywords{}
\end{abstract}
\section{Introduction}
Gravity deflects space-time. For this reason matter and energy, while bending geodesics, force other masses to move along
Keplerian trajectories. Indeed gravity, by equivalence principle and better by General Relativity, bends also massless
photons. Deflection of star's lights by Sun, nearly eighty years ago, overthrown newtonian relativity leading to the most
popular triumph of Einstein's relativity. It is the gravitational bending of the light that gave life to gravitational
microlensing (Paczy\`{n}ski \cite{pacz1}, \cite{pacz2}; Griest \cite{griest}; Alcock \cite{alcock}), one of the best ways
to see ``matter in the dark''. However gravity shapes also time rate. Indeed clocks are slowed in presence of heavy bodies
(Earth, etc.). The well known gravitational redshift is (as well as light bending) a cornerstone of General Relativity.
Time is slowed down near pulsars and totally frozen when observed, far away, near the Schwarzschild radius of a black
hole.\\
The variable gravitational redshift on a fixed period due to the source motion with respect to the gravitating body is
just ``Shapiro Phase Shift'' (Shapiro \cite{sha}). It has been predicted on early 1964 and observed, by delayed radar
echoes grazing the Sun and reflected by Mercury (1967) and Venus (1970), by Shapiro himself.  We have not (yet) a powerful
radar to inspect far away dark planets or MACHOs in our galaxy. However nature does offer very precise clocks spread far in
our galaxy: pulsars. The period of pulsars is stable, million or billion times better than our commercial clocks,
competitive with the best laboratory ones on Earth. The stability of the timing of pulsars is granted by the huge inertial
momentum of the pulsar, its consequent large angular momentum and the negligible angular momentum losses. Therefore pulsars
are the timing candles in our Shapiro delay search: their beating slowdown and speed up may inform us on the dark objects
crossing along the line-of-sight.\\
While geodesic accelerations suffer the gravitational field gradients (proportional to the inverse of square distance) at a
given point, Shapiro Phase Shift (SPS) is a path integral of the field (which decreases as the inverse of the distances):
SPS ``records'' the gravitational fields from the source to the observer and the delay is a cumulative effect. Indeed there
is both a geometrical delay (which is proportional to gravitational fields) as well as a  gravitational Shapiro delay which
is roughly the path integral of the field. The consequent Gravitational Shapiro delay is a logarithmic function of the
impact parameter and it is usually the dominant one at large impact parameter values of body encounters. This effect
increases, nearly by a factor of one hundred, the characteristic impact distance of a detectable phase delay with respect
to a microlensing event of magnification of the brightness of stars. Therefore, even the few hundreds pulsars in our galaxy
offer a large enough sample to measure a few SPS delay a year (Fargion \& Conversano \cite{goffo0}, \cite{goffo}). Their
imprint is a long-term ``anomalous'' negative contribute on the time derivative of the period of pulsars, superimposed to
the slow-down positive one. Depending on the strength of the event, the total $\dot{P}$ can be still positive but lower.
Moreover a sub-sample of pulsars close to the ecliptic projection may monitor the Solar System planets (Jupiter, Mars ...)
as well as the solar SPS, allowing a quantitative calibration and verification of the SPS delay. The same technique may be
deploied to observe far hypothetical unknown planets which may pollute at the far periphery of the Solar System. Their
eventual rare infall and near encounter with the Earth in the past might have left important geological imprints
(Fargion \& Dar \cite{dfdar}); strong tidal forces might induce Tsunami and volcanic activities  as well as life
extinction. Shapiro Phase Delay may also reveal MACHOs in our Solar System neighbors while a collective correlation of the
period derivatives may track the secret trajectories of far dark bodies in the Space.
\section{Review of the Shapiro effect on pulsars} \label{section2}
\begin{figure}
\resizebox*{\hsize}{!}{\includegraphics{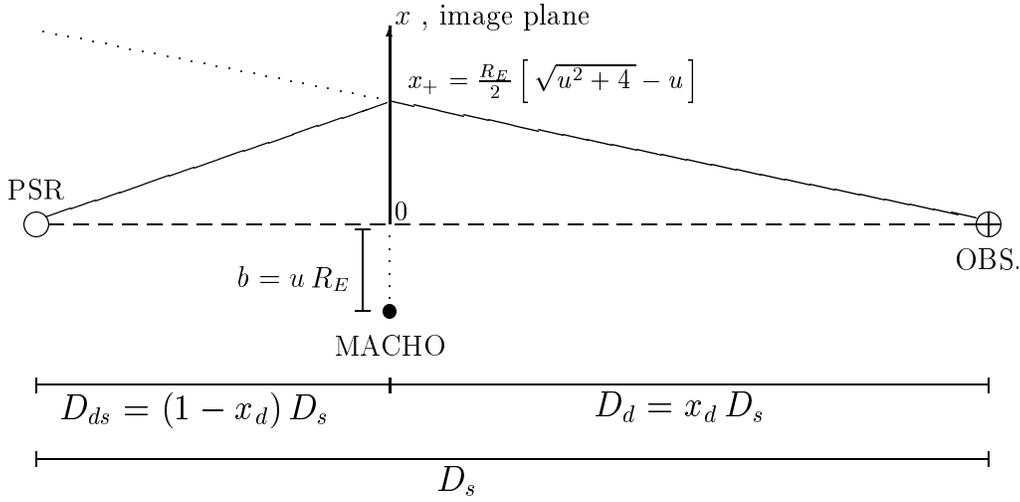}}
\caption{Relative position of source, MACHO , observer and primary image. As $u \rightarrow \infty$, $x_{+} \rightarrow 0$
         and the image corresponds to the source.$v_{\perp}$ accounts for the relative pulsar-MACHO-observer orthogonal
         motion velocity. \label{spsp}}
\end{figure}

\begin{figure}
\resizebox*{\hsize}{!}{\includegraphics{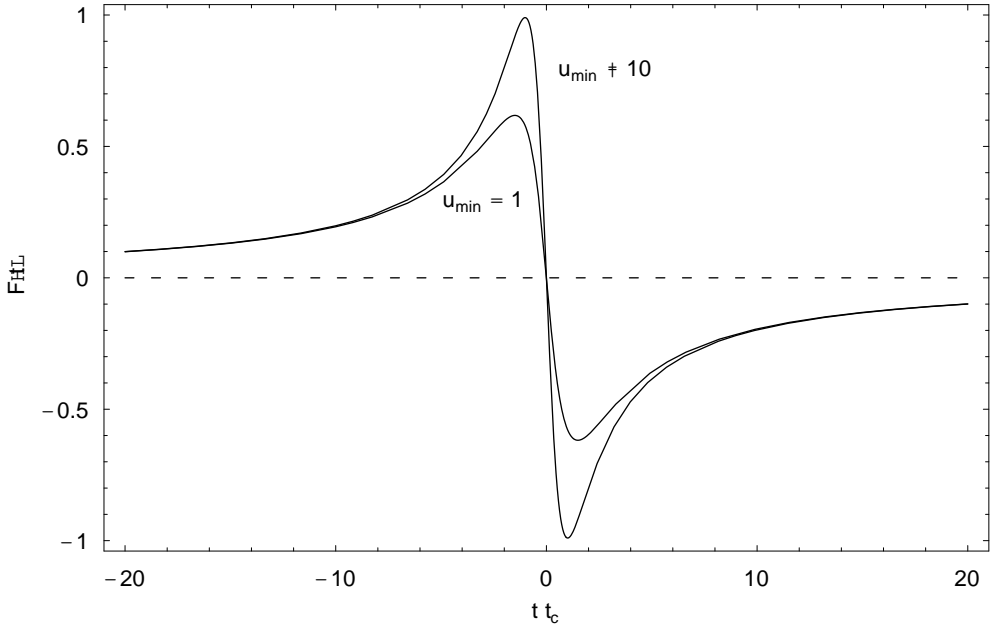}}
\caption{$F(t)$ as a function of $\,t\,/\,t_{c}\,$. \label{fderiv}}
\end{figure}

The Shapiro Phase Shift on Pulsars (SPSP) is the Shapiro effect (Shapiro \cite{sha}) which influences the time of
arrivals of the signal of the pulsars.\\
The application of the SPSP to the search of dark matter was first proposed by J. Schneider (\cite{schneiderJean90}).
Limiting to the case of strong lensing events, $<b> \simeq R_E$, where $<b>$ is the mean impact parameter estimate based on
visible matter, he constrained the observation of the phenomenon to the signals of extragalactic pulsars. Larchenkova \&
Doroshenko (\cite{larch}) probably detected an event of SPSP towards the galactic anticenter region. The authors claim the
existence of a massive black hole of $\sim 330\,M_\odot$ producing a strong lensing event. Wex, Gil \& Sendyk
(\cite{wexetal96}) investigated the possibility of identifying MACHOs in the galactic center region by means of SPSP both
for strong and weak lensing events. According to the authors the probability for observing an SPSP with lens mass between
$1$ and $20$ $M_\odot$ is quite reasonable but, for weak lensing events, it should not be possible to extract any
information on the parameters of the lens (mass, transverse velocity ...) without making assumptions about both the
intrinsic frequency $\nu$ and the intrinsic frequency derivative $\dot{\nu}$.\\
We also analyzed the SPSP effect in our previous works (Fargion \& Conversano \cite{goffo0}, \cite{goffo}).
SPSP may be responsible of the time derivative deviation and may explain the presence of a few negative pulse
derivatives among 706 known pulsars. We estimated the mean optical depth for SPSP events for weak lensing (high impact
parameters) on the basis of present models of the distribution of luminous matter in our galaxy. We found
that the SPSP may play a role in $\dot{P}$ deviations for just a $\, \sim 0.7 \, ev \,yr^{-1}\,$ event rate at a
$\,|\dot{P}| \leq 6.9 \cdot 10^{-14} \; s \, s^{-1}\,$ level for massive objects of $M \sim M_\odot$. We expect a
higher event rate for masses $M \sim 10^{-2}\,M_\odot$ $\left(\sim 7 \, ev \,yr^{-1}\right)$ but at lower intensities
of $\,|\dot{P}| \leq 6.9 \cdot 10^{-15} \; s \, s^{-1}\,$. Events toward the galactic anticenter should be rarer to observe, as a result of the rapid decrease of the density of
galactic matter with the distance from the center. Moreover, the higher the typical mass of massive objects the lower the
event rate for SPSP should be expected. On the basis of these simple probability estimates we believe the
event detected by Larchenkova \& Doroshenko be rather a weak lensing event produced by a MACHO
of mass much smaller than $\sim 330\,M_\odot$. We also showed how analogous
collective SPSP in globular clusters would also be possible.\\
The fundamental equations for SPSP can be summarized as follows.
A massive object passing by the line-of-sight of a pulsar, with transverse velocity $v_{\perp}$ at a distance
$D_{d}$ from the observer (see figure \ref{spsp}), produces a Shapiro time delay. This is made of two terms:
a first geometrical one, $\Delta t_{geo} = \frac{r_S}{4 c} \left(\, \sqrt{u^2 + 4 \;} \,-\, u\,\right)^{\,2}$,
which is the result of the angular deflection of the path and it is rapidly vanishing at large impact values:
$\left. \Delta t_{geo} \right|_{u \gg 1} = \frac{r_S}{c\,u^{\,2}}$.
The second term is the gravitational redshift due to the deflector field, $\sim \frac{G M}{c^2} \int \frac{dr}{r}$,
all along the wave trajectory and scales as the logarithm of the impact parameter $u$:
\begin{equation}\label{eqn1}
  \Delta t_{grav} \; = \; \frac{r_{S}}{c}\,\ln\left( \frac{8\,D_{s}}{r_{S}} \right) \, - \,
  \frac{2\,r_{S}}{c}\,\ln\left( \sqrt{u^2 + 4\, } + u  \right)
\end{equation}
where $D_s$ is the distance from the observer to the source and
$\;u = b \,/\, R_{E}\;$ is the dimensionless impact parameter of the MACHO
from the line-of-sight of the source, in units of the Einstein radius
\begin{equation}\label{eqn2}
  R_{E} \, = \, \sqrt{2 \, r_{S} \, \frac{D_{d} \, D_{ds}}{D_{s}} \,} \, =  4.27 \cdot 10^{13} \;
  \sqrt{\left[ \frac{M}{M_{\odot}} \right] \left[\frac{D_{s}}{4 Kpc}\right] \left[\frac{x_{d}}{0.5}\right]
  \left[ 2 - \left( \frac{x_{d}}{0.5} \right) \right]\,} \,\;\;\; cm
\end{equation}
where $\,x_{d} = D_{d} / D_{s}\,$ and $\,1 - x_{d} = D_{ds} /
D_{s}\,$ is the usual notation as found in literature ($4.84 \cdot
10^{13}\;cm\,=\,1.38 \cdot 10^{-5}\;pc\,=\,2.85\;AU$, for quick
conversion). The characteristic time of any event is
\begin{equation}\label{eqn4}
  t_{c} \, = \, \frac{R_{E}}{v_{\perp}} \; u_{min} \; \approx \; 4.51 \frac{\left[ \frac{u_{min}}{10^2} \right]}{\left[
  \frac{\beta_{\perp}}{10^{-3}} \right]} \sqrt{ \left[ \frac{M}{M_{\odot}} \right] \left[ \frac{D_{s}}{4 Kpc} \right]
  \left[\frac{x_{d}}{0.5}\right] \left[ 2 - \left( \frac{x_{d}}{0.5} \right) \right]} \;\;\;\; yrs
\end{equation}
where $u_{min}$ is the minimum impact parameter of the MACHO (see figure \ref{spsp}).
The assumption of $10^2$ as a characteristic value for $u_{min}$ offers a square geometrical
probability amplification ($\pi u^2$) to reveal a massive object, making the SPSP
on $700$ pulsars statistically comparable with the ordinary microlensing of intensity
magnification on a sample of 7 millions of stars.\\
The outstanding signature on the period of a pulsar should be a negative contribute to the
total time derivative during the second half of the SPSP event. The evolution of the
$\dot{P}$ deviation is:
\begin{equation}\label{eqn6}
  \dot{P} \; = \; \frac{d\,\Delta t_{grav}}{dt} \; = \; \frac{d\,\Delta t_{grav}}{d\,u}\;\frac{d\,u}{dt} \, = \,
  \dot{P}_{0} \, F(t)
\end{equation}
where
\begin{equation}\label{eqn7}
  \dot{P}_{0} \, = \, \frac{r_{S}\,\beta_{\perp}}{R_{E}\,u_{min}} \, = 6.92 \cdot 10^{-14} \; \frac{
  \left(\frac{\beta_{\perp}}{10^{-3}}\right) \sqrt{\frac{M}{M_{\odot}}}}{\,\left(\frac{u_{min}}{10^2}\right)
  \sqrt{\left(\frac{D_{s}}{4 Kpc}\right)\left(\frac{x_{d}}{0.5}\right)\left[2- \left(\frac{x_{d}}{0.5}\right)\right]}}
  \;\;\;\;\; s \, s^{-1}
\end{equation}
gives the strength of the event and
\begin{equation}\label{eqn8}
  F(t) \, = \, - \, 2 \, \frac{\left(\frac{t}{t_{c}}\right)}{\sqrt{\left[1 + \left( \frac{t}{t_{c}} \right)^2 \right]
  \left[1 + \left(\frac{t}{t_{c}} \right)^2 + \left( \frac{2}{u_{min}} \right)^2 \right]}} \;\;\; .
\end{equation}
its time evolution.
The function $F(t)$ is centered at the instant $t = 0$ of closest approach. It has a maximum and a symmetric negative
minimum for the values $\, \mp \, t_{c} \,$ (as shown in figure \ref{fderiv} for values $\;u_{min} = 1\;$ and
$\;u_{min} \geq 10\;$).\\
Competitive gravitational delays in residual periods of the pulsars may occur due to Doppler shifts induced by
gravitational deflection of both unbounded objects (planets, stars, black holes, ...) and binary bounded objects near the
pulsar (the former being a step-like residual signal in the period of the pulsar while the latter being a periodic
sinousoidal one). In this context the SPSP leads to a recognizable, episodic, timely bell-shaped period variation. SPSP
period derivative should be added to the average positive derivative ($10^{-14}\;s\,s^{-1}$) related to the intrinsic
pulsar spin-down. Our first inspection (Fargion \& Conversano \cite{goffo0}, \cite{goffo}) into the catalogue of 558
pulsars (Taylor et al. \cite{catalogue}) found out some ``anomalous'' negative period derivatives. In table \ref{table1}
we show the comparison between the two editions of the catalogues of the known pulsars (Taylor et al. \cite{catalogue},
\cite{catalog}), with the epoch of each measurement. We note that the pulsar B0021-72D, which had no measured value in the
older catalogue, has a negative period derivative with low absolute value. The pulsar B0021-72C, which resides in the same
globular cluster, has different values of $\dot{P}$ even if the epoch of measurement is the same (notwithstanding this
still negative). We believe the newer value of $\dot{P}$ be consistent, being similar to the value of B0021-72D and, maybe,
being evidence of a collective SPSP within the globular cluster. Particular attention must be dedicated to the pulsar
B1813-26. It does not belong to
any globular cluster nor to any binary system but increases its negative $\dot{P} = (-3 \pm 3)\,10^{-16}\;s\,s^{-1}$ to a
positive $\dot{P} = (+6.65 \pm 0.07)\,10^{-17}\;s\,s^{-1}$ after a time interval of $\sim 17.5\,yr$. This may indicate the
second phase of an SPSP event, the total $\dot{P}$ getting again positive values, as well as just an improvement of the
precision of the measure.\\
The reported values of $\dot{P}$ of the other pulsars in table \ref{table1} are equal in both the catalogues. On the basis
of these data we cannot produce a more precise statistic on the time development of the $\dot{P}$, in this sense only an
on-line frequently updated database should be useful.\\
\section{Shapiro Delay in dark planet search close to the Solar System}
\begin{figure}
\resizebox*{\hsize}{!}{\includegraphics{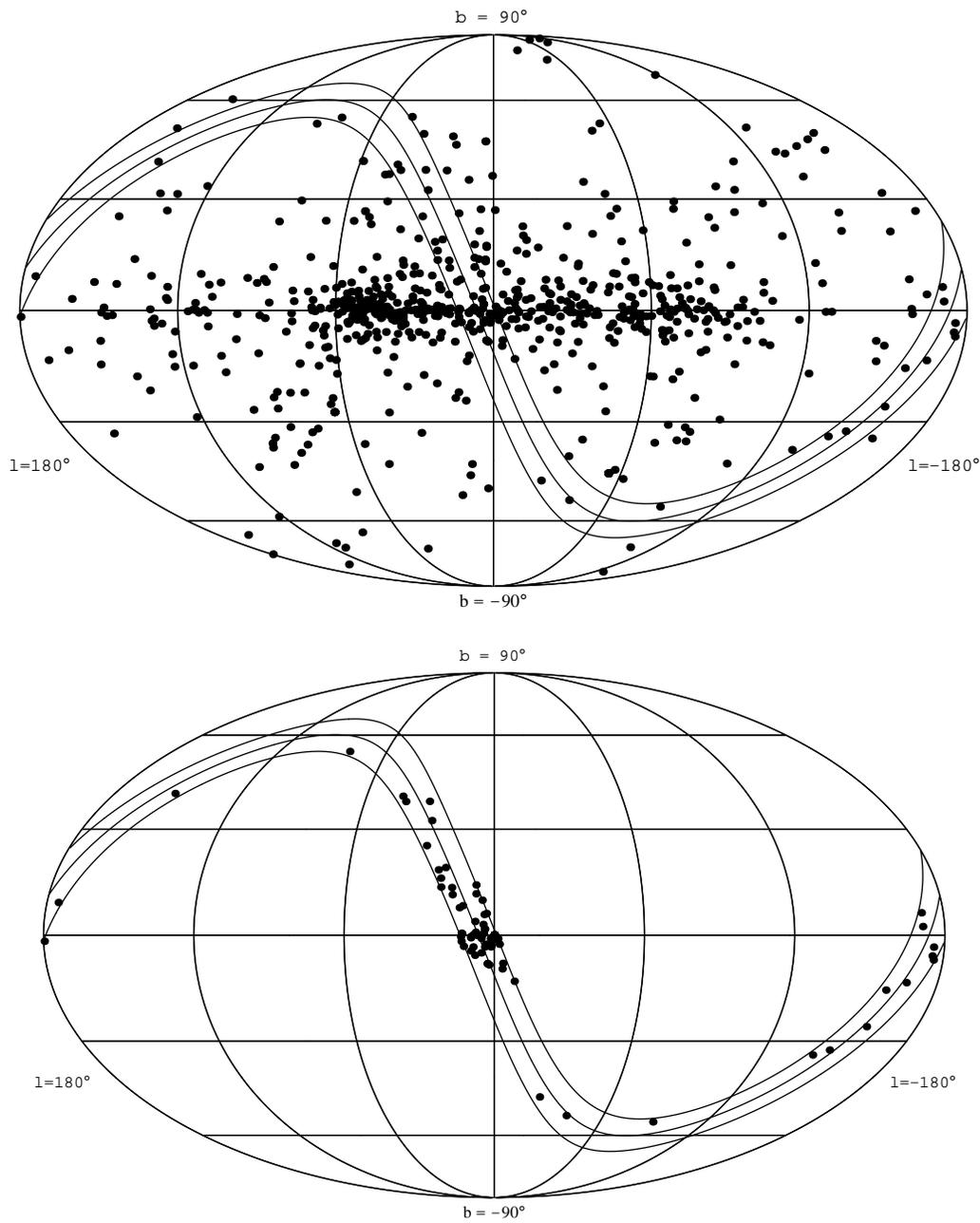}}
\caption{The upper figure shows the positions of all know pulsars (\cite{catalog}). The lower figure shows the position of
         the pulsars laying within an angular radius of $6.38^\circ$ from the ecliptic. \label{pulsarArray}}
\end{figure}

\begin{figure}
\resizebox*{\hsize}{!}{\includegraphics{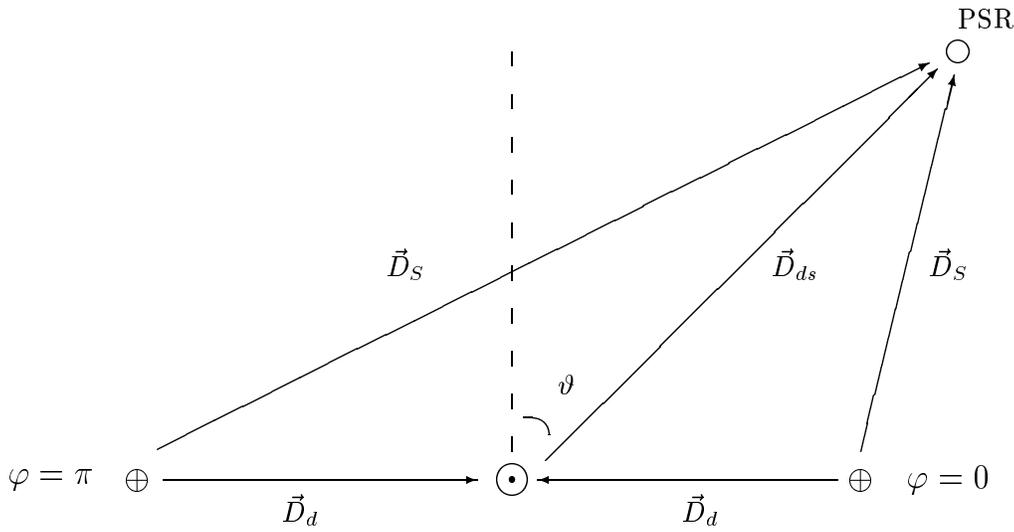}}
\caption{Relative position of the pulsar, Sun and Earth. $\varphi$ is the ecliptic longitude and $\varphi = 0$, while
         $\pi/2 - \theta$ the ecliptic latitude. \label{positionsun}}
\end{figure}

Many sources of SPSP can affect the time derivative of the period of the pulsars: local companions in binary systems,
encounters of MACHOs along the trajectory of the signal (Paczy\`{n}ski \cite{pacz1}, \cite{pacz2}; Griest \cite{griest};
Alcock \cite{alcock}), gravitational waves, collective SPSP in globular clusters (Fargion \& Conversano \cite{goffo}).
Moreover any massive object in Solar System can, in principle, be a source of SPSP. In this section we investigate on
the possibility to use the SPSP effect to discover dark planets and any other massive object close to our Solar System.
In the case of planets bounded to our Solar System we expect their orbits to lay within low ecliptic latitudes. This
defines a a preferring zone of observation close to the ecliptic. We can choose the array of pulsars laying along the
ecliptic trajectory within an angular radius of $\pm 6.38^\circ$ (so we can include the pulsars of both the galactic
center and anticenter). Figure \ref{pulsarArray} and table \ref{table2} show this subsample. We count 73 pulsars
which should be continuously monitored.\\
There are two main contributes of $\dot{P}$-deviation which affect any observed pulsar:
a Doppler effect $\dot{D}_s \,/\, c$ (as a result of both the relative motion of the pulsar and the revolution of the
Earth) and a Shapiro effect (as a result of the gravitational action of the deflector on the
optical path of the signal of the pulsar). \\
The total time delay due to the two main contributes is:
\begin{equation}\label{eqn9}
  \Delta \; t \; = \Delta \; t_{Doppler} \; + \; \Delta \; t_{Grav}
\end{equation}
We take our attention to the gravitational one. For nearby deflectors the gravitational time delay is given by the
three-dimensional vectorial formula:
\begin{equation}\label{eqn10}
  \Delta \; t_{Grav} \;=\; \frac{r_S}{c} \; \ln \left[ \; \frac{ \, D_s \,-\, D_d \, \cos \psi \,+\,D_{ds}\;}{D_d \,
  \left(\, 1 \,-\, \cos \psi \,\right) } \;\right] \;\;,
\end{equation}
where we adopted the same notation used in section \ref{section2}. $\vec{D}_d$, $\vec{D}_{ds}$ and $\,\vec{D}_s =
\vec{D}_d + \vec{D}_{ds}$ are, respectively, the position vectors of the deflector relative to the observer on Earth, the
pulsar relative to the deflector and the pulsar relative to the observer; $D_d = |\vec{D}_d|$, $D_{ds} = |\vec{D}_{ds}|$ and
$D_s = | \vec{D}_d + \vec{D}_{ds} |$ are the modules and $\psi = \widehat{\vec{D}_s \vec{D}_d}$ (figure \ref{positionsun}).
When $\psi \rightarrow 0$, $\cos \psi \sim - 1$ and $D_{ds} \,\simeq \, D_{s} \,-\, D_{d}$.
Consequently, after expanding the denominator
in equation \ref{eqn10}, we find $1 - \cos \psi \sim ( \widehat{\vec{D}_s \, \vec{D}_d} )^{\,2} /\, 2 \simeq
(\,b\,+\,x_{+}\,)^{\,2} \,/\, ( 2\,{D_{d}}^{2} ) \,=\, {R_{E}}^{2}\, (\sqrt{u^{\,2}+4}+u)^{\,2} / (8\,{D_{d}}^{2})$
(figures \ref{spsp} and \ref{positionsun}) and equation \ref{eqn10} tends to equation \ref{eqn1}. We neglect
the contribute of the geometrical deviation due to the fact that we are interested mainly in weak lensing (large $u$).
The residual period time derivative can be found by time differentiating equation \ref{eqn10}.\\
As the dark planet search involves deflector masses much lower than stellar ones, the resulting $\dot{P}$-deviations are
much lower than what expected in galactic dark matter search. Therefore a comprehensive analysis of what are all the
possible sources of SPSP in our Solar System that can produce SPSP $\dot{P}$-deviations is needed. \\
The removal of all of these effects, together with the improvement of the threshold of sensitivity, is important
for the SPSP technique in dark planet search to have success in Solar System. Here we propose two ways to achieve this:
({\it i}) the extraction of the frequency component corresponding to annual modulations on the timing;
({\it ii}) the cross-correlation of the signals coming from an array of pulsars laying along the projection of ecliptic
toward the galaxy.
\subsection{Annual $R_E$ modulation}
\begin{figure}
\resizebox*{\hsize}{!}{\includegraphics{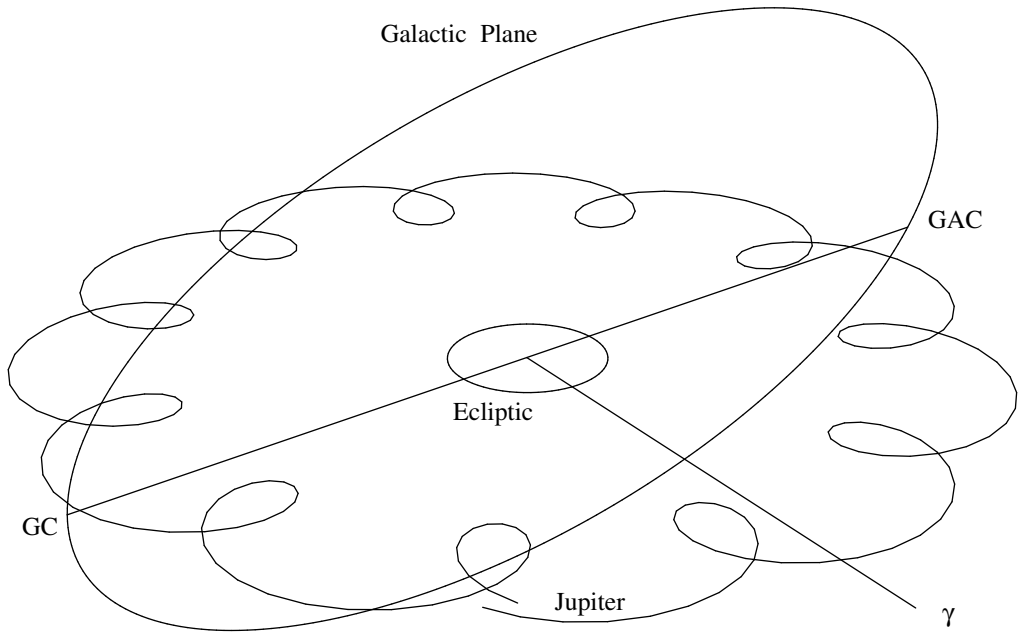}}
\caption{Geocentric orbit of the Sun, Jupiter and the projection of the galactic plane. The line labeled as $\gamma$
         indicates the vernal point direction, while the labels GC and GAC indicate the directions toward the galactic
         centre and anti-centre respectively. The orbital trajectories are calculated for a period of twelve years (UT)
         starting from 1998 January 1. \label{orbit}}
\end{figure}

The low-intensity $\dot{P}$-deviation scenario in Solar System is quite complex.
The Sun and all the known planets produce $\dot{P}$-deviations higher then or comparable to those expected in dark planet
search. Even the revolution of the Moon around the Earth and the rotation of the Earth itself produce both Doppler and SPSP
$\dot{P}$-deviations. The evaluation of the time evolution of all of these contributes is important to ensure both a
calibration and the removal of known additive disturbs in using SPSP for dark planet search. All of these
$\dot{P}$-deviations are periodic. Planets in Solar System produce periodic SPSP
$\dot{P}$-deviations with their own periods of revolution, the Sun an annual one, the Moon a monthly one and the rotation of
the Earth a daily one. The revolution of the Earth produces also an annual Doppler $\dot{P}$-deviation
(the mutual attraction of known planets in Solar System, which perturbs Earth trajectory, should be included in this term),
while the rotation of the Earth a Doppler daily one. \\
The variation of the mutual distances among the planets causes the Einstein radius $R_E$ to change.
If we refer the observations to a geocentric observer, the variation is modulated by the annual revolution of Earth.
The relative variation of the Einstein radius is:
\begin{equation}\label{RErelvar}
  \frac{\Delta R_E}{R_E} \;\simeq\; \frac{1}{2}\,\frac{\Delta D_d}{D_d}
  \,+\,\frac{1}{2}\,\frac{\Delta D_{ds}}{D_{ds}}
\end{equation}
where we do not consider the term $\Delta D_s \,/\, D_s $, being this negligible. We can distinguish three cases:
\begin{enumerate}
  \item $D_d \simeq D_{ds} \simeq D_s$. This is the most probable case, where the deflector is located nearly half-way from
        the observer and the source. The Einstein radius takes the general form of equation \ref{eqn2}.
        Relative variations of few astronomical units do not affect significantly the
        signal coming from a pulsar located at distances $D_s \sim Kpc$. In these cases planets orbiting in distant
        stellar systems should not give detectable modulations due to the orbit of the planet.
  \item $D_{ds} \ll D_d \simeq D_s$. In this case the deflector is close to the pulsar. The Einstein radius depends on the
        deflector-source distance only, $R_E \,\simeq\, \sqrt{\,2\,r_S\,D_{ds}\,}$. If the pulsar were bound in a binary
        system its signal is modulated at the frequency of rotation when $\Delta D_{ds}$ gets comparable with $D_{ds}$.
  \item $D_d \ll D_{ds} \simeq D_s$. This is the case when the deflector is very close to our Solar System.
        The Einstein radius takes the form: $R_E \,\simeq\, \sqrt{\,2\,r_S\,D_{d}\,}$, depending only on the deflector
        distance. If the deflector were a dark planet bounded to our Solar System, relative variations
        of few astronomical units should be significant. The revolution of Earth itself should infer to the signal
        an annual modulation as a result of the cyclic variation of $D_d$. As an example, figure \ref{orbit} shows
        the typical geocentric orbit of Jupiter, together with the regular orbit of the Sun. The mutual Earth-Jupiter
        distance, $D_d$, varies periodically nearly twelve times during a whole revolution of Jupiter. The period of this
        variation is exactly one year. Therefore
        we expect that the $\dot{P}$-deviation have two periodic components of modulation: a long-term one (with a period
        of nearly twelve years), as a result of the revolution of Jupiter, and a short-term one (with a period of one
        year), as a result of the revolution of Earth. Any massive object passing close to our
        Solar System should infer to the time arrivals of the pulsar a short-term $R_E$ modulation at a period of one year
        and, if bounded to Solar System, a long-term $R_E$ modulation at its own period of revolution.
        .
\end{enumerate}
The last item gives an answer to how a close SPSP event can be recognized from a far one.
-) First, the annual frequency component must be extracted from the timing of the pulsar. From this component both the
Doppler effect and the solar SPSP, mainly, must be subtracted. Now, the resulting signal is still affected by
the short-term $R_E$-modulated components of some other known planets in our Solar System. The key point is now to
recognize which planets. \\
-) As the SPSP modulations due to the other known planets can be easily calculated, one should check if long-term
frequency components are present in the residual timing of the pulsar. If a long-term component at a known
frequency, say, that of Jupiter, is present, the corresponding short-term $R_E$ component must also be subtracted from the
annual component. Therefore, as any known planet produces its own characteristic long-term component, one can, in
principle, recognize the presence of SPSP contributes from known planets and subtract them.\\
-) Any residual annual frequency component, free from both Doppler and known-SPSP effects, if still present, could reveal
a massive nearby object.\\
Now, what is the sensitivity? We can compare the change in Shapiro time delay, as a result
of a change in $D_d$ ($\delta_d \Delta t_{grav}$) with the $\dot{P}$ deviation. The procedure to calculate $\dot{P}$
is (Fargion \& Conversano \cite{goffo}):
\begin{equation}\label{REmodulationEq1}
  \dot{P} \;\simeq\; \frac{\partial\,\Delta t_{grav}}{\partial\,u} \, \frac{\partial\,u}{\partial\,t}
\end{equation}
where
\begin{equation}\label{REmodulationEq2}
  u(t) \;=\; u_{min}\,\sqrt{\,1\,+\,\left(\frac{t}{t_c}\right)^2\;} \;\;,
\end{equation}
while $\delta_d \Delta t_{grav}$ is:
\begin{equation}\label{REmodulationEq3}
  \delta_d \Delta t_{grav} \;\simeq\; \frac{\partial\,\Delta t_{grav}}{\partial\,u} \;
                                    \frac{\partial\,u}{\partial\,D_d} \; \Delta D_d \;\;.
\end{equation}
Comparing equations \ref{REmodulationEq1} and \ref{REmodulationEq3} and approximating $R_E \simeq \sqrt{2 r_S D_d}$,
we find $\delta_d \Delta t_{grav}$ as a function of $\dot{P}$ and the relative variation $\Delta D_d / D_d$:
\begin{equation}\label{REmodulationEq4}
  \delta_d \Delta t_{grav} \;\;=\;\; {\Delta t}_d^{\,0} \;\cdot\; F_{d}(t)
\end{equation}
where
\begin{equation}\label{REmodulationEq5}
  {\Delta t}_d^{\,0} \;=\; \frac{1}{2}\,t_c\,\dot{P}_0\,\frac{\Delta D_d}{D_d}\; \,=\, 1.10 \;\; \left(\frac{t_c}{1\;yr}\right)\,
                           \left(\frac{\dot{P}_0}{6.92 \cdot 10^{-14}s\,s^{-1}}\right)\,
                           \left(\frac{\Delta D_d}{D_d}\right)\;\;\; \mu s
\end{equation}
and
\begin{equation}\label{REmodulationEq6}
  F_{d}(t) \;=\; \left(\frac{t}{t_c}\right) \; F(t) \;\; .
\end{equation}
\subsection{Multiple SPSP on an array of pulsars}
\begin{figure}
\resizebox*{\hsize}{!}{\includegraphics{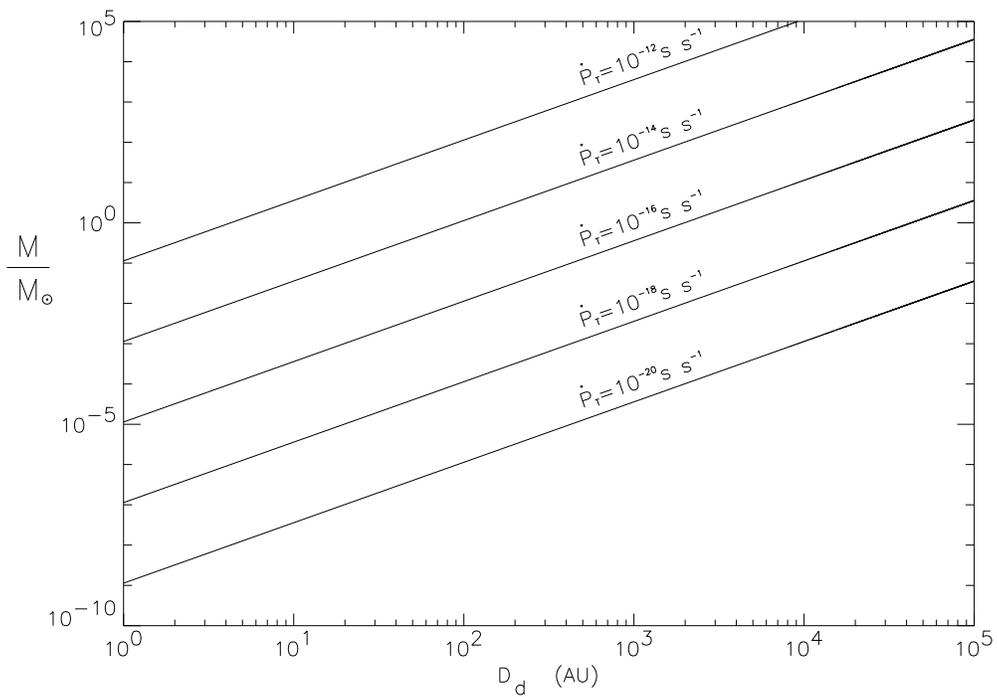}}
\caption{Mass of the deflector required to produce a $\dot{P}_T$-level deviation on the $\left| \dot{P} \right|$ of the
         pulsar. The angular impact parameter is the angular width in latitude of the subsample of pulsars shown
         in figure \ref{pulsarArray} and table \ref{table2}. The plot is a function of the distance $D_d$ of the
         hypothetical planet from the center of the Solar System. \label{ArrayThresholdFigure}}
\end{figure}

Any massive object either orbiting around or passing by close to our Solar System may slow down the signal of more than
one pulsar at the same time. In fact, the projection of its impact parameter subtends an angle wider than that subtended
in the case of a far deflector.
In this case we can use the whole array of known pulsars both to increase the detectability threshold and to trace
the trajectory of the object via a collective statistics. With what threshold of detectability we can hope to achieve this?
We can choose the maximum acceptable angular impact parameter as, say, the latitude width of the subsample in table
\ref{table2}
($\Delta b = 2 \cdot 6.38^\circ$). With this choice we can calculate what $\dot{P}$-deviation a planet passing by one edge of the subsample
generates on a pulsar laying on the other edge. The linear impact parameter $R_E u_{min}$ becomes
$2 \, D_d \, \tan (6.38^\circ)$ and the $\dot{P}$-deviation (equation \ref{eqn7}) produced by the planet is:
\begin{equation}\label{eqArray1}
   \dot{P}_T = \frac{r_{S}\,\beta_{\perp}}{2 D_d \, \tan (6.38^\circ)} \;\;\;.
\end{equation}
We can replace the dependence from $\beta_{\perp}$ with the distance $D_d$ of the deflector from the observer:
\begin{equation}\label{eqArray2}
  \beta_{\perp} = \frac{\beta_{\oplus}}{\sqrt{\frac{r}{r_{\oplus}} \;}} \simeq
  \frac{\beta_{\oplus}}{\sqrt{\frac{D_d}{r_{\oplus}} \;}}
\end{equation}
where the distance $r$ of the deflector from the Sun has been approximated to $D_d$ for large distances,
$r_{\oplus} = 1\,AU$ and $\beta_{\oplus} = 9.96 \cdot 10^{-5}$. In this case equation \ref{eqArray1} becomes:
\begin{equation}\label{eqArray3}
  \dot{P}_T \simeq \frac{\beta_{\oplus}}{2\,r_{\oplus} \, \tan (6.38^\circ)}
  \frac{r_{S}}{\left( \frac{D_d}{r_{\oplus}}\right)^{\frac{3}{2}}} = 8.8 \cdot 10^{-22} \, \left(
  \frac{M}{10^{-10}\,M_\odot}\right) \; \left( \frac{D_d}{1\,AU}\right)^{\,-\,\frac{3}{2}} \; s \; s^{-1} \;\;\;.
\end{equation}
Equation \ref{eqArray3} can be used to determine the minimum mass of a planet that can perturb the signal of a pulsar
(with almost the maximum impact parameter allowable within the two edges) of the subsample at a $\dot{P}_T$-level
deviation (see figure \ref{ArrayThresholdFigure}). Then, the characteristic time of the $\dot{P}_T$-level deviation is:
\begin{equation}\label{eqArray4}
  t_{c_T} \;=\; \frac{R_E u_{min}}{\beta_{\perp} c} \;=\; \frac{2 D_d \tan (6.38^\circ)}{\beta_{\perp} c} \;\simeq\;
  1.12 \cdot 10^6 \; \left( \frac{D_d}{1\,AU}\right)^{\frac{3}{2}} \; s \;\;\;.
\end{equation}
Figure \ref{ArrayThresholdFigure} shows the mass $M$ of the deflector plotted as a function of its distance $D_d$ for many
threshold levels $\dot{P}_T$. Only planets with mass greater than $\sim 10^{-3}\;M_\odot$ can produce the actually-observed
average $\dot{P}$-deviation, $\dot{P}_T \sim 10^{\,-14}\;s\,s^{\,-1}$, across the two edges of the subsample.
To detect mini-planets ($M \geq 10^{-10}\;M_\odot$), a $\dot{P}_T \leq 10^{-20}\;s\,s^{\,-1}$ threshold of sensitivity
should be reached.

\section{Solar-induced gravitational time delay}
\begin{figure}
\resizebox*{\hsize}{!}{\includegraphics{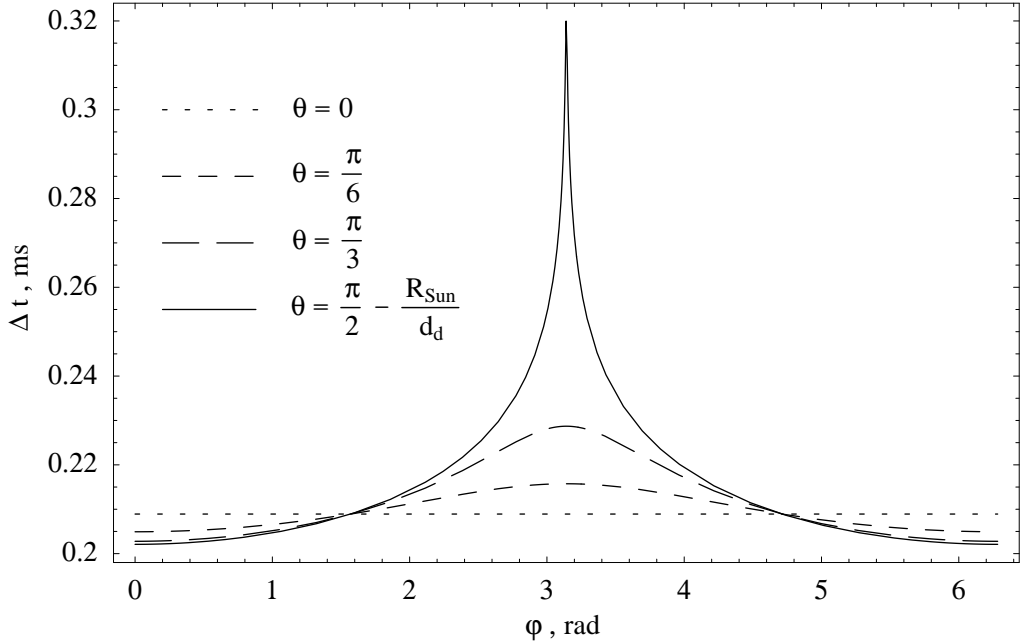}}
\caption{Gravitational time delay produced by the Sun on the timing of a pulsar laying at a mean distance $d_S = 4 Kpc$,
         plotted for four values of the orientation of the ecliptic. When $\vartheta=\frac{\pi}{2}-\frac{R_{Sun}}{D_d}$
         the position vector of the pulsar, $\vec{D_s}$, grazes the surface of the Sun. \label{deltasun}}
\end{figure}

\begin{figure}
\resizebox*{\hsize}{!}{\includegraphics{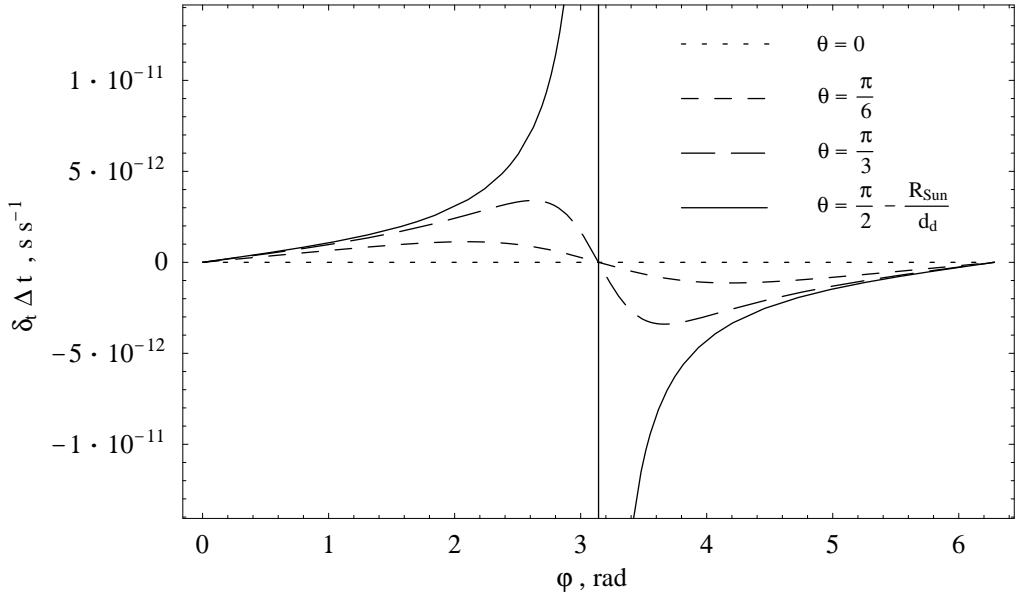}}
\caption{Residual period time derivative on the timing of a pulsar as a result of the gravitational field of the Sun for
         the same ecliptic latitudes and $D_s$ chosen for the plot shown in figure \ref{deltasun}. The complete plot for
         the maximum $\dot{P}$ deviation is shown in figure \ref{ppointsun}. \label{ppointsunzoom}}
\end{figure}

\begin{figure}
\resizebox*{\hsize}{!}{\includegraphics{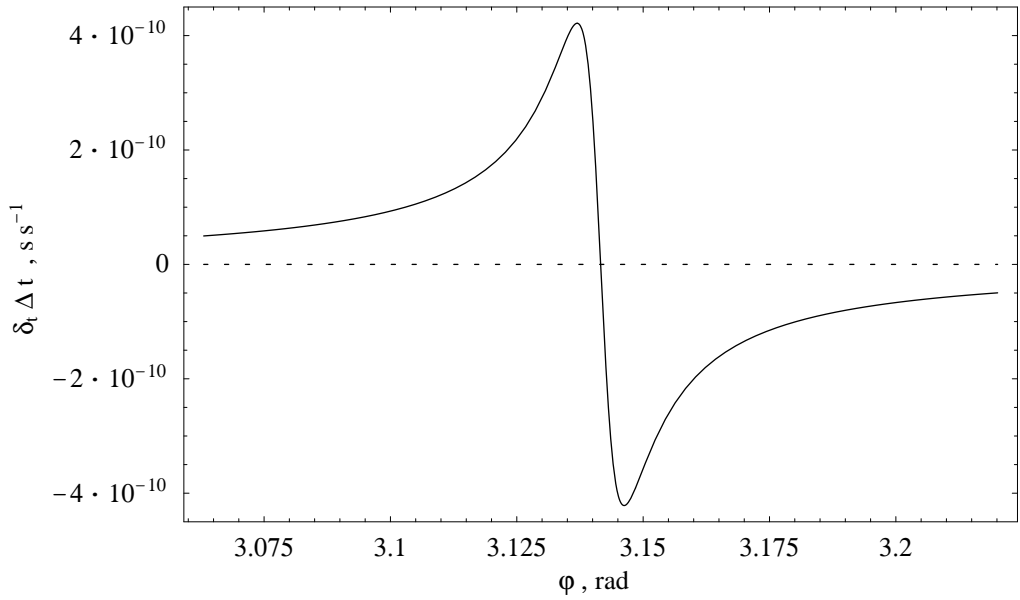}}
\caption{Total plot of the maximum residual period time derivative as a result of the gravitational field of the Sun for
         the case $\vartheta=\frac{\pi}{2}-\frac{R_{Sun}}{D_d}$.  \label{ppointsun}}
\end{figure}

\begin{figure}
\resizebox*{\hsize}{!}{\includegraphics{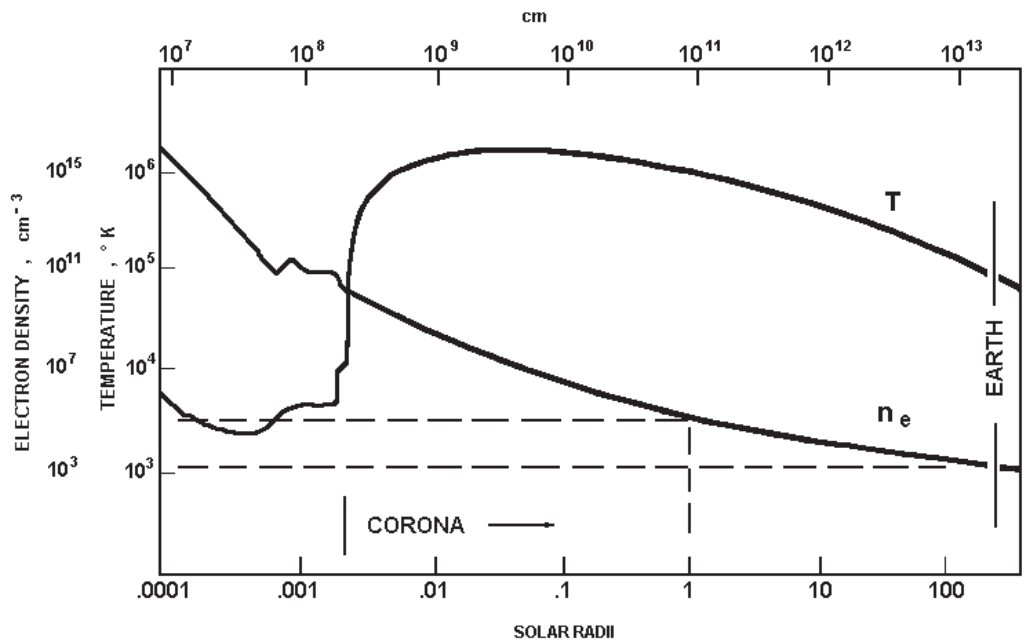}}
\caption{Electron density, $n_e$, from the solar surface. \label{electron}}
\end{figure}

The detection of dark planets close to Solar System is expected at low ecliptic latitudes. The SPSP deviation of the known
planets and the Sun must be calculated and subtracted from the timing of the observed pulsars.
The SPSP contribute from the Sun does not
suffer the $R_E$ annual modulation because, with except for lower-order eccentricity effects, the observer-deflector
distance $D_d$ does not vary. Figure \ref{deltasun} shows the Shapiro time delay $\Delta \; t_{\odot_{Grav}}$ induced by
the Sun on the timing of a pulsar at an average distance of $4\,Kpc$ (Taylor et al. \cite{catalogue}). The four curves
correspond to four different values of the ecliptic latitude of the pulsar. The maximum Shapiro time delay is found when
the position vector of the pulsar, $\vec{D_s}$, grazes the surface of the Sun and decreases at high ecliptic latitudes.
Figures \ref{ppointsunzoom} and \ref{ppointsun} show the corresponding residual time derivatives of the period, $\dot{P}$.\\
The formulation used in section \ref{section2} is valid for impact parameters $b = u_{min}\,R_E\ll D_d$ and it is
suitable to evaluate the maximum SPSP effect of the Sun. Taking characteristic values of $D_{d} \approx 1 \; AU$ and
$\beta_{\perp} \approx 10^{-4}$ we find:
\begin{equation}\label{re_sun}
  R_E \simeq \sqrt{\;2 r_S \, D_d \;} \approx 2.97 \cdot 10^{ 9} \; \sqrt{\,\left( \frac{M}{M_\odot} \right)\,
             \left( \frac{D_d}{1 \; AU} \right)\;}   \;\; cm \;,
\end{equation}
\begin{equation}\label{tc_sun}
  t_c = \frac{R_E \; u_{min}}{v_\perp} \approx 9.91 \cdot 10^4 \; \left( \frac{u_{min}}{10^2} \right)
        \left( \frac{\beta_\perp}{10^{-4}} \right)^{-1} \; \sqrt{\, \left( \frac{M}{M_\odot} \right) \,
        \left( \frac{D_d}{1 \; AU} \right) \;} \;\; s \;,
\end{equation}
\begin{equation}\label{dotP0_sun}
  \dot{P}_0 = \frac{r_S\,\beta_\perp}{R_E \; u_{min}} \approx 9.94 \cdot 10^{-11}\; \left(\frac{u_{min}}{10^2}\right)^{-1}
              \left( \frac{\beta_\perp}{10^{-4}} \right) \; \left( \frac{M}{M_\odot} \right)^{\frac{1}{2}} \,
              \left( \frac{D_d}{1 \; AU} \right)^{-\frac{1}{2}} \;\;\; s \; s^{\,-1}  \;\;\;\; s\,s^{-1}.
\end{equation}
The SPSP effect is huge ($|\dot{P}| \simeq 4.2 \cdot 10^{-10}\;s\,s^{\,-1}$) when $b = u_{min}\,R_E = R_\odot$
($u_{min} = 23.41$). The characteristic time of the maximum SPSP effect is $t_c \simeq 0.27\,day$.\\
In order to avoid any additional and confusing refractive index delay (due to solar plasma even at large impact parameters)
we must compare the two effects by comparing the perturbations, $\Delta n_{grav}$ and $\Delta n_p$, induced on the
refraction index. The gravitational refraction index is:
\begin{equation}\label{refract_grav}
  n_{grav} \simeq 1 - \frac{2}{c^2} \; U = 1 + 4.22 \cdot 10^{-6} \left( \frac{M \;}{M_\odot} \right)
                  \left( \frac{r \;}{R_\odot} \right)^{-1}\;\;,
\end{equation}
while the plasma refraction index:
\begin{equation}\label{refract_plasma}
  n_p (\nu) = \sqrt{1 - \left( \frac{\nu_p}{\nu} \right)^2} \simeq 1 - \frac{1}{2} \left( \frac{\nu_p }{\nu} \right)^2 =
              1 - 4.03 \cdot 10^{-6} \left( \frac{n_e }{ 10^5 \, cm^{-3}}\right) \left(\frac{\nu }{1\,GHz}\right)^{-2},
\end{equation}
where $\nu_p = \sqrt{\,\left( e^2 n \right) \,/\, \left( \pi m \right) \; } = 8.98 \cdot 10^3 \sqrt{\, n_e\;} \;\; Hz$.
For electrons, we have
\begin{equation}\label{refract_comp}
  \frac{\Delta n_{grav}}{\Delta n_p} \simeq 1.05 \; \left(\frac{\nu }{ 1 \, GHz} \right)^2
            \left( \frac{n_e }{ 10^5 \, cm^{-3}} \right)^{-1} \left( \frac{r \,}{R_\odot} \right)^{-1} \;\;.
\end{equation}
The density of electrons decreases with the distance from the solar surface (see figure \ref{electron}). At a distance of
one solar radius from the surface of the Sun ($r = 2\,R_\odot$, $u_{min} \simeq 46.8$) $n_e \sim 10^{\,5} \; cm^{-3}$
and therefore $\Delta n_{grav}/\Delta n_p \simeq \; 0.53 \, \left[\nu / \left(1\,GHz\right) \right]^2$. Close to Earth
$n_e \, \sim \, 2 \cdot 10^{\,3} \;\; cm^{-3}$ and $r = 1\;AU$ and, again, $\Delta n_{grav} / \Delta n_p \sim 0.49
\, \left[\nu / \left(1\,GHz\right) \right]^2$. From these values we deduce that, near Sun, pulsars might be better observed
at high radio frequencies ($\nu \geq 1.4 GHz$), which include the few known optical $(\,X,\,\gamma\,)$ pulsars. In this
high frequency range the SPSP effect of the Sun is dominant.

\section{Jupiter-induced gravitational time delay}
\begin{figure}
\resizebox*{\hsize}{!}{\includegraphics{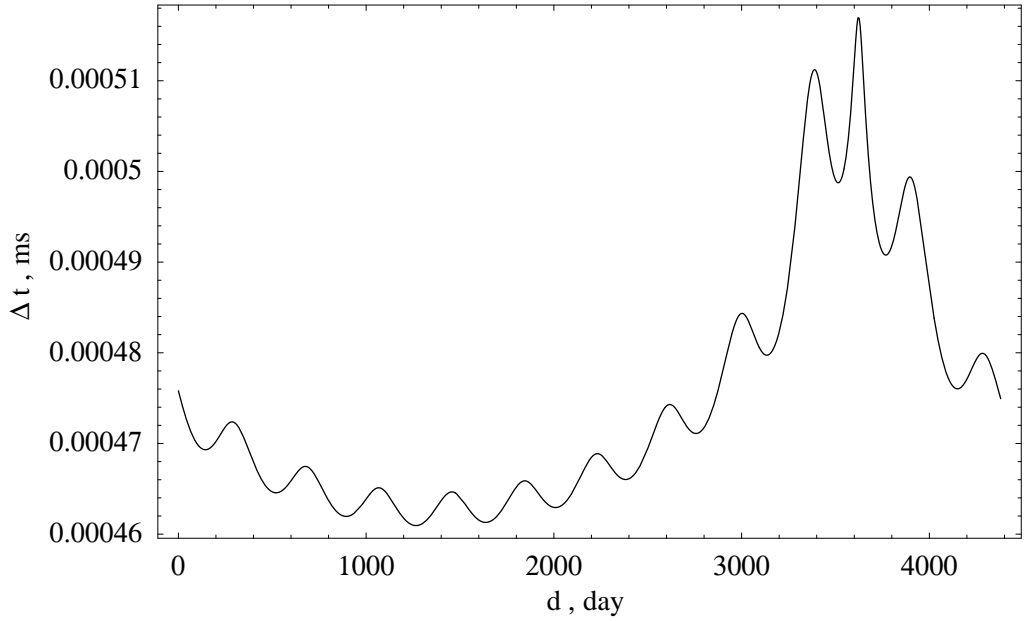}}
\caption{Time delay as a result of the gravitational effect of Jupiter on the signal of an hypothetical pulsar laying at a
distance of $4\,Kpc$ in the galactic centre direction. \label{deltaTjupGeo}}
\end{figure}

\begin{figure}
\resizebox*{\hsize}{!}{\includegraphics{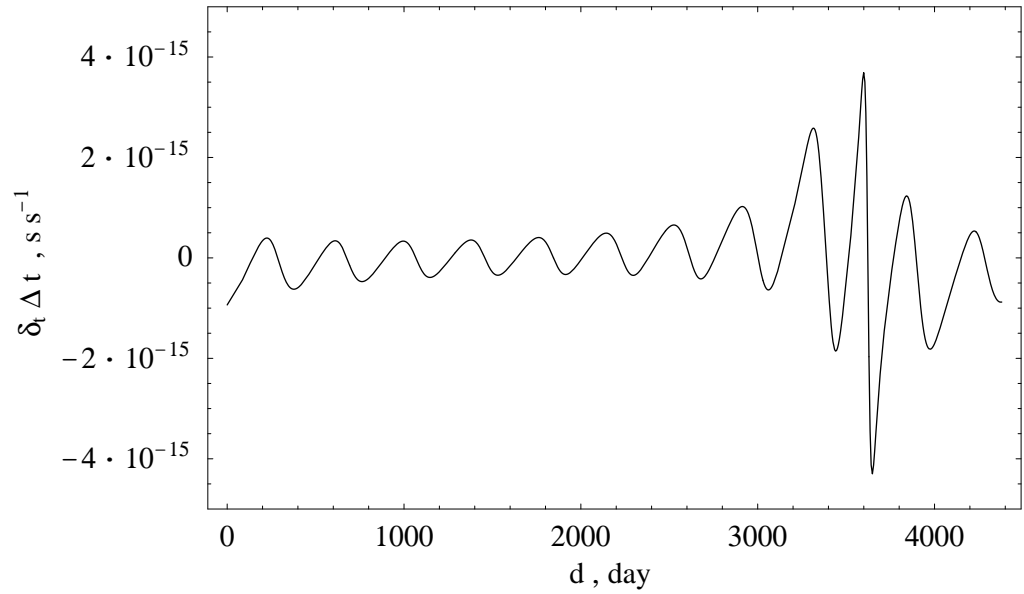}}
\caption{Residual period time derivative for the plot in figure \ref{deltaTjupGeo}.
 \label{deltaTderivjupGeo}}
\end{figure}

Jupiter is the most massive planet in our Solar System that can produce SPSP. Its geocentric orbit is a
representative example of the annual $R_E$ modulation. Figure \ref{orbit} shows the geocentric orbit of Jupiter during a
twelve-year revolution. Because the distance of Jupiter from the Sun is about $5.2\,AU$, the geocentric distance $D_d$
varies periodically from $4.2\,AU$ to $6.2\,AU$ ($\Delta/D_d \simeq 0.2$), nearly twelve times during a revolution of
Jupiter.\\
The evaluation of the maximum SPSP effect Jupiter can produce can be made following the same procedure used for
the case of the Sun. We find:
\begin{equation}\label{re_jupiter}
  R_E \simeq \sqrt{\;2 r_S \, D_d \;} \approx 2.14 \cdot 10^{\,8} \; \sqrt{\; \left( \frac{M}{10^{-3} M_\odot} \right) \,
             \left( \frac{D_d}{5.2 \; AU} \right) \;} \;\; cm \;,
\end{equation}
\begin{equation}\label{tc_jupiter}
  t_c = \frac{R_E \; u_{min}}{v_\perp} \approx 7.15 \cdot 10^3 \; \left( \frac{u_{min}}{10^2} \right)
        \left( \frac{\beta_\perp}{10^{-4}} \right)^{-1} \; \sqrt{\; \left( \frac{M}{10^{-3} M_\odot} \right) \,
        \left( \frac{D_d}{5.2 \; AU} \right) \;} \;\; s \;,
\end{equation}
\begin{equation}\label{dotP0_jupiter}
  \dot{P}_0 = \frac{r_S\,\beta_\perp}{R_E \, u_{min}} \approx 1.38 \cdot 10^{-12} \left( \frac{u_{min}}{10^2}
              \right)^{-1} \left( \frac{\beta_\perp}{10^{-4}} \right) \left( \frac{M}{10^{-3} M_\odot}
              \right)^{\frac{1}{2}} \, \left( \frac{D_d}{5.2 \, AU} \right)^{-\frac{1}{2}} s\,s^{-1}.
\end{equation}
The equatorial radius of Jupiter is $R_J = 7.14 \cdot 10^9\,cm$, its mass is $M_J \simeq 0.78 \cdot 10^{-3}\,M_\odot$
and the average speed of revolution $\beta_\perp \simeq 4.36 \cdot 10^{-5}$. Therefore, the average Einstein radius of
Jupiter is $R_E \simeq 1.89 \cdot 10^8 \; cm$ ($u_{min} = 37.82$) and the maximum SPSP reaches a level of
$|\dot{P}| \simeq 4.47 \cdot 10^{-14} \; s\,s^{-1}$ for a characteristic time of $t_c \simeq 173 \;s$.\\
We also performed a simulation of the realistic SPSP produced by Jupiter on a distant pulsar. The pulsar is located at an
average distance of $4\,Kpc$ in the galactic centre direction.
Figure \ref{deltaTjupGeo} shows the gravitational time delay during a twelve-year cycle of revolution of Jupiter around
the Sun, as seen from a geocentric observer. We can distinguish twelve peaks of gravitational time delay the magnitude of
which increases as the position of Jupiter tends to the galactic centre direction. We can recognize the classic bell-shaped
curve of a SPSP with a one-year modulation superimposed. Figure \ref{deltaTderivjupGeo} shows the corresponding
residual time derivative of the period. The effect is not the maximum possible because the minimum impact
parameter of Jupiter with respect to the source is large compared with the Einstein radius (a little less than $1\,AU$).

\section{Mars-induced gravitational time delay}

\begin{figure}
\resizebox*{\hsize}{!}{\includegraphics{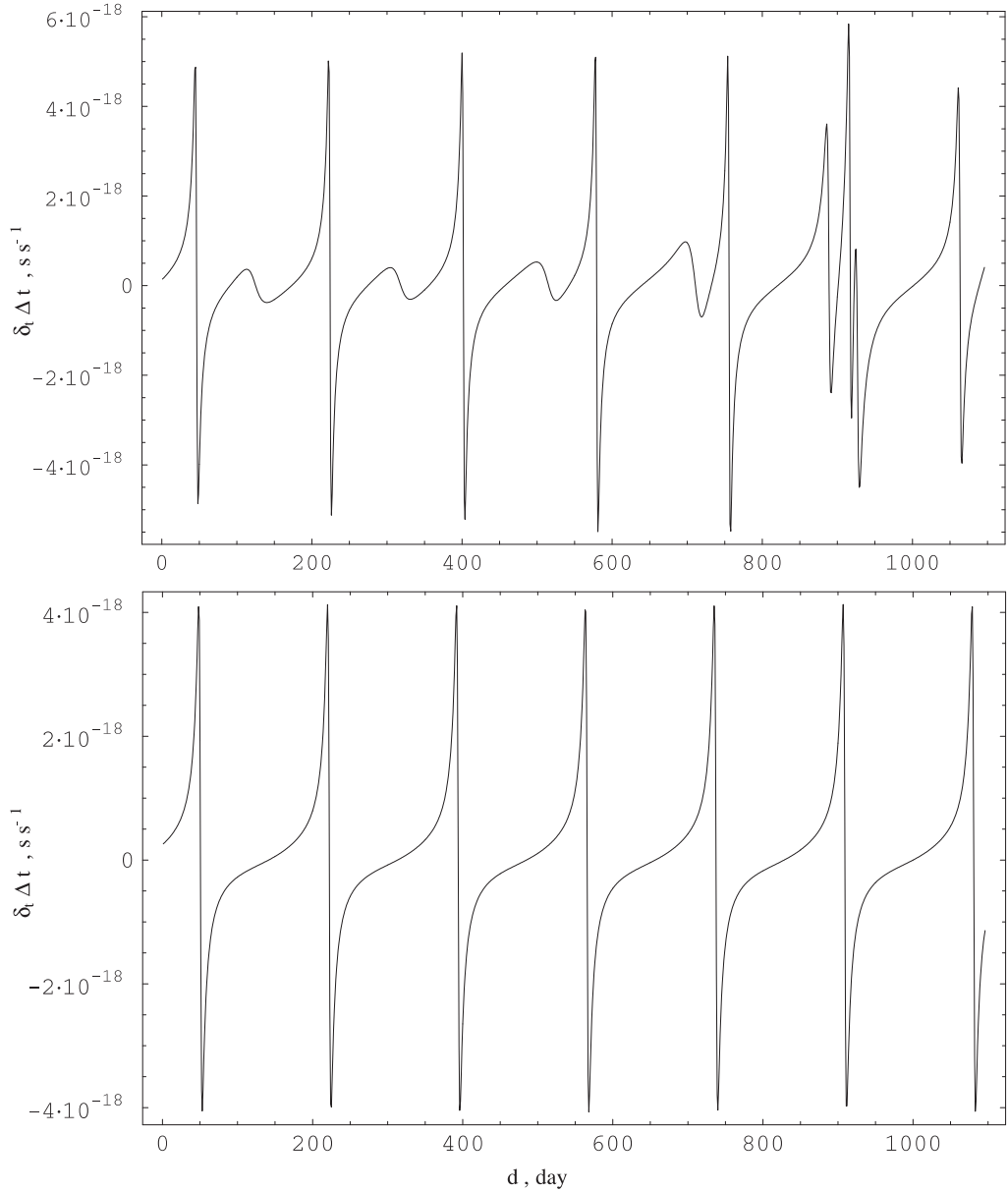}}
\caption{Residual time derivative for a pulsar laying in the galactic centre direction at a distance of $4\,Kpc$, affected
by the gravitational field of Mars, as observed by a geocentric observer (upper figure). The lower figure shows the same
event with the modulation removed. The simulation covers the same twelve-year time interval taken in the case of Jupiter.
 \label{Marsderiv}}
\end{figure}

When we take planets with orbits closer to the Sun then that of Jupiter, the period of revolution of gets comparable to
one year and the modulation gets more strong and complicated in its time evolution. The case of Mars is shown in figure
\ref{Marsderiv}. Again, the pulsar is located in the galactic centre direction. The
Einstein radius of Mars is much smaller than that of Jupiter while the minimum impact parameter reached in the simulation
is practically the same. This implies a much larger $u_{min}$ and, so, a weaker event.\\
Again, the maximum effect can be deduced from the equations:
\begin{equation}\label{re_mars}
  R_E \approx 1.15 \cdot 10^{\,6} \; \sqrt{\; \left( \frac{M}{10^{-7} M_\odot} \right) \,
             \left( \frac{D_d}{1.5 \; AU} \right) \;} \;\;\;\; cm \;,
\end{equation}
\begin{equation}\label{tc_mars}
  t_c \approx 38.40 \; \left( \frac{u_{min}}{10^2} \right)
        \left( \frac{\beta_\perp}{10^{-4}} \right)^{-1} \; \sqrt{\; \left( \frac{M}{10^{-7} M_\odot} \right) \,
        \left( \frac{D_d}{1.5 \; AU} \right) \;} \;\;\;\; s \;,
\end{equation}
\begin{equation}\label{dotP0_mars}
  \dot{P}_0 \approx 2.57 \cdot 10^{-14} \; \left( \frac{u_{min}}{10^2}
              \right)^{-1} \left( \frac{\beta_\perp}{10^{-4}} \right) \; \left( \frac{M}{10^{-7} M_\odot}
              \right)^{\frac{1}{2}} \, \left( \frac{D_d}{1.5 \; AU} \right)^{-\frac{1}{2}} \;\;\;\; s\,s^{-1}.
\end{equation}
The mass of Mars is $M_M \simeq 2.63 \cdot 10^{-7}\,M_\odot$, the radius $R_M = 3.39 \cdot 10^8\,cm$
($u_{min} = 294.78$) and the average speed of revolution $\beta_\perp \simeq 8.05 \cdot 10^{-5}$.
The maximum SPSP reaches a level of $|\dot{P}| \simeq 1.14 \cdot 10^{-14} \; s\,s^{-1}$ for a characteristic time of
$t_c \simeq 228 \;s$.

\section{Effects due to Earth rotation and Moon revolution}

\begin{figure}
\resizebox*{5cm}{!}{\includegraphics{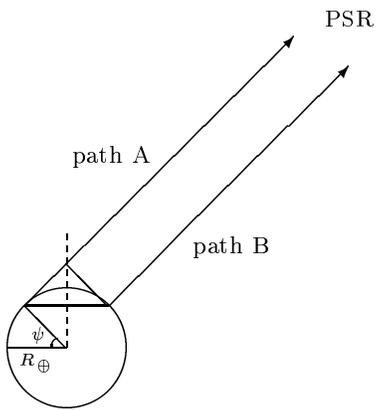}}
\caption{Path variation as a result of the rotation of Earth.
 \label{earthrot}}
\end{figure}

We now check what kind of gravitational disturb both the moon and
the rotation of the Earth should produce on the period time
derivative of the pulsar. In the case of the Moon we can easily
evaluate the maximum effect of the Shapiro effect from our usual
formulas, when the trajectory of the signal grazes the surface of
the moon. Setting \( D_{d}=3.84\cdot 10^{10}\: cm \) as the mean
Moon-Earth distance , \( R_{moon}=1.74\cdot 10^{8}\: cm \) as the
radius of the Moon and \( r_{S_{\: moon}}=r_{S_{\, \oplus }}/81.3
\) as the Schwarzschild radius of the Moon the corresponding
Einstein radius is \( R_{E_{\, moon}}=2.89\cdot 10^{\, 4}\: cm \).
Taking \( R_{moon} \) as the impact parameter \(u_{min}\:
R_{E_{\,moon}}\) we find
\begin{equation}\label{moonppoint}
\dot{P}_{0_{\, moon}}\: =\frac{2\, r_{S_{\, \oplus }}\, \left\langle \beta _{\perp }
\right\rangle }{R_{moon}}=4.29\cdot 10^{\, -16}\; \; s\; s^{\, -1}\; \; \;,
\end{equation}
which is quite a small effect.\\
In addition we should take into account the gravitational perturbation produced
by the rotation of the Earth. Let us take a simple limiting case.
The pulsar is visible by the observer during the whole
rotation and, at some instant of time, the trajectory of the signal is
orthogonal to the Earth radius (see figure \ref{earthrot}). In
this simple case the maximum time delay during a rotation of Earth is:
\begin{equation}\label{earthrotppoint}
\Delta t_{rot} \,\simeq\, \frac{\, r_{S_\oplus}}{\, c \;} \; \ln
\left[ \; \sin (2\,\psi ) + \sqrt{1 + {{\sin (2\,\psi )}^2}} \;
\right] \;\;,
\end{equation}
where $\psi$ is the latitude of the observer. Suppose it be
$2\,\pi\,/\,3$, then $\Delta t_{rot} \,\simeq\, 0.78 \cdot r_{S_\oplus} \,/\, c \,=\, 2.31 \cdot 10^{\,-11}\;\;
s$ and the mean residual time period derivative would be
$\langle \,\dot{P}\,\rangle \,\simeq\, \Delta t_{rot} \,/\, (0.5\,day) \,=\, 5.35 \cdot 10^{\,-16} \;\;s\; s^{\, -1}$.

\section{Conclusions}
The Shapiro Phase might be able to discover not only dark matter but, once calibrated, it may even lead to new mini
planets discovers on our ecliptic plane. Their presence might be of relevance by tidal disturbances during rare encounters
on Earth past and future history and life evolution (\cite{dfdar}). Finally recent EGRET data (\cite{dixon}) on diffused
Gev galactic-Halo might find an answer, among the others (\cite{df3}), by ($\sim pc$) hydrogen molecular clouds interacting
by proton (Gev) cosmic rays (\cite{depaolis}). Microlenses at those large radii, (by Kirchoff theorem) are inefficient.
Therefore the Shapiro Phase Delay might be the unique probe to verify such an evanescent dark matter barionic
candidature.\\
When taking in account the existence of celestial objects trapped by our Solar System we can view a one-year modulation on
the SPSP produced by them as an outstanding unique feature. Being produced by celestial objects laying only near the
geocentric observer this modulation can be used as a discrimination method. Among the possible methods of enhancing the
identification of some low-mass celestial objects within our Solar System we considered the observation of more than one
pulsar at a time within a subset of pulsars near the ecliptic.
A random combined statistic on many pulsars would deteriorate the predicted one-year modulation, because different
modulations on many pulsars would interfere destructively. On the contrary a phase matching of the modulation may
greatly increase by coherent interference the sensitivity, leading to the discover of the vectorial motion of the
``dark'' object. In any case, as a first point of view on this matter, we can say that the combination of both the
methods could be taken into act. A one-year modulated warning system would be useful for low-mass objects which can
affect the residual timing of a pulsar at intensities well within the present observational detectability, while a
combined statistic on many pulsars would be useful for the discovery of very low-mass celestial objects which would
produce an SPSP of very low intensity. In view of a possible need of the discover of dangerous impact trajectories of
asteroids, and in view of a need of monitoring incoming mini-planets, the cross-time correlation of the periods of a net
of pulsars may offer a unique tool able to see in the dark.

%
\newpage
\begin{table}[t]
\caption{Pulsars with negative $\dot{P}$. Comparison between the
$\dot{P}$ in the two catalogues: (Taylor et al. \cite{catalogue})
fourth and fifth columns and (Taylor et al. \cite{catalog}), sixth
and seventh columns. \label{table1} }
\begin{tabular}{@{}lllllll}
\hline
       &          &           & \multicolumn{2}{c}{$\dot{P}$ in the old catalogue} & \multicolumn{2}{c}{$\dot{P}$ in the new catalogue} \\
pulsar & Environment & $P$ ($ms$)& $\dot{P}$ ($\;s\,s^{-1}$)&
Epoch (MJD) & $\dot{P}$ ($\;s\,s^{-1}$)& Epoch (MJD)  \\
\hline
%
 B0021-72C & C    & $5.8  $ & $(-4 \pm 5)\,10^{-17}$ & $47858.521$ & $(-4.98 \pm 0.01)\,10^{-20}$ & 47858.5  \\
 B0021-72D & C    & $5.4  $ & $                        $ & $47858.521$ & $(-2.8 \pm 0.2)\,10^{-21}$ & 48040.7  \\
 B1744-24A & B , C& $11.6 $ & $(-1.9 \pm 0.2)\,10^{-20}$ & $48270.0  $ & $(-1.9 \pm 0.2)\,10^{-20}$ & 48270.0  \\
 B1813-26  &      & $592.9$ & $(-3 \pm 3)\,10^{-16}$ & $42004.1  $ & $(+6.65 \pm 0.07)\,10^{-17}$ & 48382.0  \\
 B2127+11D & C    & $4.8  $ & $(-1.075 \pm 0.012)\,10^{-17}$ & $47632.52 $ & $(-1.075 \pm 0.012)\,10^{-17}$ & 47632.52 \\
 B2127+11A & C    & $110.6$ & $(-2.107 \pm .003)\,10^{-17}$ & $47632.52 $ & $(-2.107 \pm .003)\,10^{-17}$ & 47632.52 \\
 \hline
\end{tabular}
\medskip \\
In the second right column the notations mean : C = globular
\mbox{cluster ,} B =  binary pulsar \vspace{7cm}
\end{table}
\begin{table}
  \centering
  \caption{Table of pulsars laying within an angular radius of $\sim 6.38^\circ$ close to the ecliptic.}
  \label{table2}
\begin{tabular}{llrrrr}
   \hline
 Pulsar B  & Pulsar J   & l (deg)   & b (deg)&  P (s) &  \multicolumn{1}{c}{$ \dot{P} \;\; (s\;s^{\,-1})$}  \\
   \hline
 0254-53   & 0255-5304  &   -90.13  & -55.31 & 0.448  &    $(3.06 \pm 0.03)\;10^{-17}$ \\
 0447-12   & 0450-1248  &  -148.92  & -32.62 & 0.438  &    $(1.033 \pm 0.008)\;10^{-16}$ \\
 0450-18   & 0452-1759  &  -142.92  & -34.08 & 0.549  &    $(5.7564 \pm 0.0004)\;10^{-15}$ \\
           & 0459-0210  &  -158.60  & -25.75 & 1.133  &    $(0.00 \pm 0.00) \;\;\;\;\;\;\;\;\;\;$ \\
 0523+11   & 0525+1115  &  -167.30  & -13.24 & 0.354  &    $(7.362 \pm 0.008)\;10^{-17}$ \\
 0525+21   & 0528+2200  &  -176.14  &  -6.89 & 3.746  &    $(4.00321 \pm 0.00006)\;10^{-14}$ \\
           & 0533+04    &  -159.91  & -15.32 & 0.963  &    $(1.50 \pm 0.40)\;10^{-16}$ \\
 0531+21   & 0534+2200  &  -175.44  &  -5.78 & 0.033  &    $(4.209599 \pm 0.000002)\;10^{-13}$ \\
           & 0538+2817  &   179.71  &  -1.68 & 0.143  &    $(3.665 \pm 0.005)\;10^{-15}$ \\
 0540+23   & 0543+2329  &  -175.63  &  -3.31 & 0.246  &    $(1.542378 \pm 0.000006)\;10^{-14}$ \\
 0609+37   & 0612+3721  &   175.45  &   9.09 & 0.298  &    $(5.89 \pm 0.01)\;10^{-17}$ \\
 0611+22   & 0614+2229  &  -171.20  &   2.39 & 0.335  &    $(5.9630 \pm 0.0005)\;10^{-14}$ \\
 0626+24   & 0629+2415  &  -171.18  &   6.22 & 0.477  &    $(1.99705 \pm 0.00019)\;10^{-15}$ \\
 0917+63   & 0921+6254  &   151.43  &  40.72 & 1.568  &    $(3.6081 \pm 0.0004)\;10^{-15}$ \\
           & 1518+4904  &    80.80  &  54.28 & 0.041  &    $(0.00 \pm 0.00) \;\;\;\;\;\;\;\;\;\;$ \\
           & 1627+1419  &    30.02  &  38.31 & 0.491  &    $(3.93 \pm 0.03)\;10^{-16}$ \\
 1633+24   & 1635+2418  &    42.99  &  39.88 & 0.491  &    $(1.194 \pm 0.002)\;10^{-16}$ \\
           & 1640+2224  &    41.05  &  38.27 & 0.003  &    $(2.9 \pm 0.2)\;10^{-21}$ \\
           & 1645+1012  &    27.71  &  32.54 & 0.411  &    $(8.13 \pm 0.02)\;10^{-17}$ \\
 1709-15   & 1711-1509  &     7.41  &  14.01 & 0.869  &    $(1.104 \pm 0.001)\;10^{-15}$ \\
           & 1713+0747  &    28.75  &  25.22 & 0.005  &    $(8.52 \pm 0.02)\;10^{-21}$ \\
 1718-02   & 1720-0212  &    20.13  &  18.93 & 0.478  &    $(8.7 \pm 0.2)\;10^{-17}$ \\
 1717-16   & 1720-1633  &     7.37  &  11.53 & 1.566  &    $(5.82 \pm 0.02)\;10^{-15}$ \\
 1718-19   & 1721-1936  &     4.86  &   9.74 & 1.004  &    $(1.59 \pm 0.02)\;10^{-15}$ \\
 1726-00   & 1728-0007  &    23.02  &  18.28 & 0.386  &    $(1.124 \pm 0.004)\;10^{-15}$ \\
           & 1730-2304  &     3.13  &   6.02 & 0.008  &    $(1.9 \pm 1.9)\;10^{-20}$ \\
 1730-22   & 1733-2228  &     4.02  &   5.74 & 0.872  &    $(4.21 \pm 0.09)\;10^{-17}$ \\
 1732-02   & 1734-0212  &    21.90  &  15.92 & 0.839  &    $(4.2 \pm 0.2)\;10^{-16}$ \\
 1732-07   & 1735-0724  &    17.27  &  13.28 & 0.419  &    $(1.21496 \pm 0.00009)\;10^{-15}$ \\
 1738-08   & 1741-0840  &    16.95  &  11.30 & 2.043  &    $(2.274 \pm 0.004)\;10^{-15}$ \\
 1740-03   & 1743-0337  &    21.67  &  13.41 & 0.445  &    $(3.17 \pm 0.14)\;10^{-15}$ \\
 1740-13   & 1743-1351  &    12.69  &   8.20 & 0.405  &    $(4.81 \pm 0.06)\;10^{-16}$ \\%
           & 1744-2334  &     4.47  &   2.97 & 1.683  &    $(0.00 \pm 0.00) \;\;\;\;\;\;\;\;\;\;$ \\
 1742-30   & 1745-3040  &    -1.44  &  -0.96 & 0.367  &    $(1.06649 \pm 0.00001)\;10^{-14}$ \\
 1745-12   & 1748-1300  &    14.01  &   7.65 & 0.394  &    $(1.212 \pm 0.002)\;10^{-15}$ \\
 1745-20   & 1748-2021  &     7.72  &   3.80 & 0.289  &    $(4.00 \pm 0.01)\;10^{-16}$ \\
 1744-24A  & 1748-2446A &     3.83  &   1.69 & 0.012  &    $(-1.9 \pm 0.2)\;10^{-20}$ \\
 1744-24B  & 1748-2446B &     3.84  &   1.69 & 0.443  &    $(0.00 \pm 0.00) \;\;\;\;\;\;\;\;\;\;$ \\
  \hline
\end{tabular}\end{table}
\begin{table}
  \centering
\begin{tabular}{llrrrr}
   \hline
               \multicolumn{6}{c}{{\bf Table 2.} Continued} \\
   \hline
 1746-30   & 1749-3002  &    -0.54  &  -1.24 & 0.610  &    $(7.871 \pm 0.002)\;10^{-15}$ \\
 1747-31   & 1750-3157  &    -2.01  &  -2.51 & 0.910  &    $(1.97 \pm 0.02)\;10^{-16}$ \\
 1749-28   & 1752-2806  &     1.53  &  -0.96 & 0.563  &    $(8.1394 \pm 0.0008)\;10^{-15}$ \\
 1750-24   & 1753-2502  &     4.25  &   0.50 & 0.528  &    $(1.413 \pm 0.002)\;10^{-14}$ \\
 1753-24   & 1756-2435  &     5.02  &   0.04 & 0.671  &    $(2.8500 \pm 0.0001)\;10^{-14}$ \\
 1754-24   & 1757-2421  &     5.30  &   0.01 & 0.234  &    $(1.30 \pm 0.03)\;10^{-14}$ \\
 1756-22   & 1759-2205  &     7.46  &   0.80 & 0.461  &    $(1.08 \pm 0.14)\;10^{-14}$ \\
           & 1759-2922  &     1.19  &  -2.87 & 0.574  &    $(0.00 \pm 0.00) \;\;\;\;\;\;\;\;\;\;$ \\
 1757-23   & 1800-2343  &     6.13  &  -0.12 & 1.031  &    $(0.00 \pm 0.00) \;\;\;\;\;\;\;\;\;\;$ \\
 1758-23   & 1801-2306  &     6.81  &  -0.07 & 0.416  &    $(1.12983 \pm 0.00003)\;10^{-13}$ \\
 1757-24   & 1801-2451  &     5.25  &  -0.88 & 0.125  &    $(1.2790 \pm 0.0001)\;10^{-13}$ \\
 1758-29   & 1801-2920  &     1.43  &  -3.24 & 1.082  &    $(3.292 \pm 0.004)\;10^{-15}$ \\
 1800-21   & 1803-2137  &     8.39  &   0.14 & 0.134  &    $(1.34327 \pm 0.00001)\;10^{-13}$ \\
 1800-27   & 1803-2712  &     3.49  &  -2.53 & 0.334  &    $(1.73 \pm 0.03)\;10^{-17}$ \\
           & 1804-2718  &     3.54  &  -2.82 & 0.009  &    $(0.00 \pm 0.00) \;\;\;\;\;\;\;\;\;\;$ \\
 1804-27   & 1807-2715  &     3.84  &  -3.26 & 0.828  &    $(1.225 \pm 0.006)\;10^{-14}$ \\
 1805-20   & 1808-2057  &     9.45  &  -0.39 & 0.918  &    $(1.7099 \pm 0.0009)\;10^{-14}$ \\
 1806-21   & 1809-2109  &     9.41  &  -0.71 & 0.702  &    $(3.822 \pm 0.001)\;10^{-15}$ \\
           & 1809-3547  &    -3.41  &  -7.84 & 0.860  &    $(0.00 \pm 0.00) \;\;\;\;\;\;\;\;\;\;$ \\
 1809-173  & 1812-1718  &    13.10  &   0.53 & 1.205  &    $(1.9119 \pm 0.0004)\;10^{-14}$ \\
 1809-176  & 1812-1733  &    12.90  &   0.38 & 0.538  &    $(9.72 \pm 0.08)\;10^{-16}$ \\
 1813-17   & 1816-1729  &    13.43  &  -0.42 & 0.782  &    $(7.254 \pm 0.006)\;10^{-15}$ \\
 1813-26   & 1816-2649  &     5.21  &  -4.90 & 0.593  &    $(6.65 \pm 0.07)\;10^{-17}$ \\
 1814-23   & 1817-2312  &     8.48  &  -3.28 & 0.626  &    $(0.00 \pm 0.00) \;\;\;\;\;\;\;\;\;\;$ \\
 1813-36   & 1817-3618  &    -3.19  &  -9.37 & 0.387  &    $(2.0465 \pm 0.0014)\;10^{-15}$ \\
 1817-18   & 1820-1818  &    13.20  &  -1.72 & 0.310  &    $(9.41 \pm 0.03)\;10^{-17}$ \\
 1819-22   & 1822-2256  &     9.34  &  -4.37 & 1.874  &    $(1.353 \pm 0.003)\;10^{-15}$ \\
           & 1822-4210  &    -8.12  & -12.85 & 0.457  &    $(0.00 \pm 0.00) \;\;\;\;\;\;\;\;\;\;$ \\
 1820-30A  & 1823-3021A &     2.78  &  -7.91 & 0.005  &    $(3.385 \pm 0.001)\;10^{-18}$ \\
 1820-30B  & 1823-3021B &     2.78  &  -7.91 & 0.379  &    $(3.15 \pm 0.03)\;10^{-17}$ \\
 1820-31   & 1823-3106  &     2.12  &  -8.27 & 0.284  &    $(2.92294 \pm 0.00018)\;10^{-15}$ \\
 1821-19   & 1824-1945  &    12.27  &  -3.10 & 0.189  &    $(5.2248 \pm 0.0013)\;10^{-15}$ \\
 1821-24   & 1824-2452  &     7.79  &  -5.57 & 0.003  &    $(1.61845 \pm 0.00005)\;10^{-18}$ \\
 2151-56   & 2155-5641  &   -22.95  & -47.05 & 1.374  &    $(4.23 \pm 0.10)\;10^{-15}$ \\
 2321-61   & 2324-6054  &   -39.57  & -53.17 & 2.348  &    $(2.60 \pm 0.11)\;10^{-15}$ \\ %
  \hline
\end{tabular}\end{table}
\end{document}